\documentclass[twocolumn,showpacs,preprintnumbers,amsmath,amssymb]{revtex4}

\usepackage{graphicx}
\usepackage{dcolumn}
\usepackage{bm}

\begin{document}


\title{Lyapunov  indices
with two nearby trajectories in a curved spacetime}

\author{Xin Wu$^{1,2}$}
\email{xwu@ncu.edu.cn}
\author{Tian-Yi Huang$^2$}
\email{tyhuang@nju.edu.cn}
\author{Hong Zhang$^2$}
\affiliation{1. Department of Physics, Nanchang University,
Nanchang 330047, China \\ 2. Department of Astronomy, Nanjing
University, Nanjing 210093, China}


\begin{abstract}
We compare three methods for computing invariant Lyapunov
exponents (LEs) in general relativity. They involve the geodesic
deviation vector technique (M1), the two-nearby-orbits method with
projection operations and with coordinate time as an independent
variable (M2), and the two-nearby-orbits method without projection
operations and with proper time as an independent variable (M3).
An analysis indicates that M1 and M3 do not need any projection
operation. In general, the values of LEs from the three methods
are almost the same. As an advantage, M3 is simpler to use than
M2. In addition, we propose to construct the invariant fast
Lyapunov indictor (FLI) with two-nearby-trajectories and give its
algorithm in order to quickly distinguish chaos from order. Taking
a static axisymmetric spacetime as a physical model, we apply the
invariant FLIs to explore the global dynamics of phase space of
the system where regions of chaos and order are clearly
identified.
\end{abstract}

\maketitle

\section{Introduction}
Chaos is often visible in nonlinear systems. General relativity as
a nonlinear theory is potentially chaotic. Although many features
of chaos in Newtonian dynamics have been known clearly for over
forty years, their application in relativistic astrophysics began
to be widely appreciated only within the last decade or so [1].
One main interest lies in studying the difference of dynamics
between Newtonian and relativistic trajectories. Varadi \emph{et
al.} [2] noted that the general relativity effects are small for
the outer planets but not negligible. Additionally, Wanex [3]
revealed the chaotic amplification effect in the relativistic
restricted three-body problem (namely, the  ideal
spacecraft-earth-moon orbital system ). In particular, several
authors found chaos in two relativistic systems including the two
fixed black holes [4] and the Schwarzschild black hole plus a
dipolar shell [5,6], which does not appear in their corresponding
Newtonian counterparts at all. Another important problem that has
been developed in recent years is to question whether spinning
compact binaries [7-15] as promising sources of gravitational
waves exhibit chaotic behavior because the gravitational-wave
detection can not succeed when chaos is present. In practice, all
the examples are attributed to the geodesic or non-geodesic motion
of particles in a given gravitational field. On the other hand,
the time evolution of the gravitational field itself, such as the
mixmaster cosmology [4,16,17], is also of great interest.

In general, the methods for quantifying the ordered or chaotic
nature of orbits in general relativity follow those widely applied
in classical physics. Let us recall a fraction of these classical
methods briefly. Poincar\'{e}'s surface of section is one of the
most common qualitative tools in the analysis of conservative
Newtonian dynamical systems of not more than 2 degrees of freedom
or with 3 degrees of freedom and axial symmetry. However, this
technique is difficult to describe a higher dimensional phase
space. Certainly, the principal Lyapunov exponent (LE), as a
measure of the average exponential deviation of two nearby orbits,
is frequently used. There are two different algorithms for the
calculation of LE: the variational method and the two-particle one
[18] (see Sec. II). The algorithm of LE  is applicable to a phase
space with any dimension, but a high dimension would cause
extremely expensive computation to get a reliable value of LE.
There are also other qualitative methods for multi-dimensional
systems, such as the power spectra, smaller alignment index (SALI)
[19,20] and fast Lyapunov indicators (FLIs) [21,22] etc. The power
spectra display a finite number of discrete frequencies for
regular orbits, whereas they are continuous for chaotic ones. One
of its problems is that it is ambiguous to differentiate among
complicated periodic orbits, quasi-periodic orbits and weakly
chaotic orbits. As far as the SALI method is concerned, it is
based on the difference or the sum of two normalized deviation
vectors at the same points of an orbit. If the dimension of phase
space is larger than 2, the indicator changes around non-zero
values for regular orbits, while it tends exponentially to zero
for chaotic orbits. The FLI uses the logarithm of the length of a
deviation vector, which increases following completely different
time rates for different orbits thus allowing one to distinguish
between the ordered and chaotic cases (see Sec. III for some
details). Both the SALI and the FLI are not only very fast tools
to find chaos, but also can sketch out the global structure of
phase space. Obviously it takes more CPU time for the former than
for the latter since the former requires two deviation vectors,
while the latter needs only one.

Considering the above various cases, we are mainly in favor of
applying two Lyapunov indices, the LE and the FLI, to study the
geodesic or non-geodesic motion of particles in relativistic
gravitational systems with at least two degrees of freedom. As is
well known, the classical definition of LE depends on the choice
of time and space coordinates in general relativity, because a
coordinate gauge in relativity can be arbitrarily adopted so that
time and space coordinates are not necessarily physical. In other
words, there would be different values of the classical LEs in
different coordinate systems. Even such LEs may provide wrong
information on the dynamical features of a system. For instance,
the maximum LE of a chaotic system turns out to be zero after a
logarithmic time transformation, that is to say, chaos becomes
hidden. Consequently, it is strongly desirable to develop  a
coordinate independent definition of LEs for the study of
relativistic dynamics.

In general relativity, one can still follow the two approaches to
calculate LEs as in the Newtonian case. As a counterpart of the
variational technique, a geodesic deviation vector can be obtained
from the geodesic deviation equation for a given geodesic flow in
general relativity. Using it, one can easily get a method (M1) for
the invariant LEs in the configuration space as well as in the
phase space. In addition, Imponente and Montani [17] presented an
invariant treatment by projecting a geodesic deviation vector for
the Jacobi metric on an orthogonal tetradic basis. In this way
they can succeed in gaining an insight into the dynamics of the
mixmaster cosmology. On the contrary, it is seldom to see in the
references that LEs are directly computed by use of the geodesic
deviation equation. Perhaps it is rather troublesome to derive the
complicated curvature tensor. In this sense, it is very natural to
extend the two-particle method from the classical to the
relativistic case. It is convenient to employ the method (M2)
introduced by Wu and Huang [23]. They compute gauge invariant
Lyapunov exponents by calculating the space separation between an
``observer'' and a ``neighbor'' particles that move along two
neighboring orbits in the phase space. Here it is necessary to
make use of a ``1+3" split of the  observer's spacetime and its
space projected operator. Meanwhile, Wu and Huang [23] use
coordinate time as the time variable of the equations of motion.
Besides M2, there is another invariant two-particle method (M3),
in which the integration time variables are the individual proper
times of the two particles instead of the common coordinate time
and the projection operation is not used. A method like M3 can be
seen in many references (for example, see [8,12]) where the proper
time and the Euclidian distance in the phase space are adopted,
but we prefer to utilize the Riemannian distance in the
configuration space than the Euclidian one in the phase space for
conceptual clarity.

Tancredi \emph{et al.} [24]  have compared the two approaches for
the calculation of LEs in Newtonian dynamics. It is still
necessary to do a comparison of the three methods because general
relativity is much different from the Newtonian dynamics. A
fundamental task is to select an optimal and valid algorithm to
compute LEs. Another aim of our work is to provide a simple, rapid
and efficient indicator of chaos which is independent of the
dimension of phase space and the choice of time and space
coordinates. It is assumed that this indicator not only can tell
us some information about the global motion in a complicated
multi-dimensional relativistic system, but also study the
transition from regular motion to chaos as certain physical
parameters alter.

In principle, it is possible to fulfill the two purposes in terms of
LE. Udry and Pfenniger [25] made a detailed quantitative estimate of
chaos in a series of 24 triaxial models of elliptical galaxies by
means of LEs. They claimed that an interval of about 2 Hubble times
corresponding to 66 periods or so is sufficient to get stable values
of LEs for each of one hundred randomly selected orbits in every
model. However, as Contopoulos and Barbanis [26] found, lots of
these orbits must be calculated for 12500 periods, and sometimes up
to 250000 periods for the accomplishment of reliable values of LEs.
On the basis of the values of the LEs, they displayed the structure
of phase space of the systems. In addition, Caranicolas and
Papadopoulos [27] studied the transition from regular motion to
chaos in a 2-dimensional logarithmic potential of elliptical
galaxies by observing the Poincar\'{e} surface of section as some
dynamical parameters change. Of course, it is necessary to calculate
thousands of orbits in the above cases, and each of them must be
computed for long enough time in general. Consequently, the time
required to compute such numbers is rather expensive, even would
reach the limit of the present numerical experiments. Under these
circumstances, one had better take advantage of the FLI as a quicker
and more sensitive indicator to separate chaotic orbits from regular
ones. As is mentioned above, the indicator is originally based on
the tangential vectors from the variational equations of Newtonian
dynamics [21]. Pondering the difficulty and complicacy to derive
variational equations in general relativity, we shall be interested
in developing the FLI method with the two particles approach
--- the FLI computed with two-nearby-orbits. It should be emphasized that
this approach would meet difficulty without renormalization in the
numerical calculation, different from the original approach of FLIs
[21] with the tangential vector. An algorithm of FLI in the present
paper will be provided.

This paper is organized as follows. The classical definition of LE
is reviewed briefly, and the three methods M1, M2 and M3 in
relativistic systems are introduced in Sect. II.  Meanwhile, we
offer a theoretical analysis of M1 and M3 to explain why they do
not need projection operations. In Sect. III the FLI with
two-nearby-orbits and its algorithm are presented. Setting a
static axisymmetric spacetime composed of a Schwarzchild black
hole and an octopolar shell  as a test model, we investigate the
global dynamics of this system. Finally, the summary follows in
Sec. IV.

Throughout the work we use units $c=G=1$, and take the signature
of the metric as $(-,+,+,+)$. Greek subscripts run from 0 to 3 and
Latin indexes from 1 to 3. The symbol $\circ$ denotes the
Euclidian inner product; while $|\mbox{\boldmath$a$}|$ is the
length of the vector $\mbox{\boldmath$a$}$ in the Euclidian space.
The symbol $|a|$ represents the absolute value of a scalar $a$. In
addition, $\bullet$ stands for the Riemannian inner product, and
$\|\mbox{\boldmath$\xi$}\|$ is the Riemannian norm of the vector
$\mbox{\boldmath$\xi$}$ corresponding to the tensor $\xi^{\alpha}$
or $\xi_{\alpha}$. $\mbox{\boldmath$f$}$  is a set of functions.
We specify $t$ as a time coordinate variable, and $\tau$, a proper
time variable. $D/D\tau$ denotes a covariant derivative operator
vs $\tau$.

\section{Comparisons of three methods for computing LE
in general relativity}

\subsection{LE in classical physics}

In classical physics the LE is to characterize the mean exponential
rate of divergence of trajectories surrounding a given trajectory in
the phase space. For a compact autonomous $n$-dimensional system
\begin{equation}
\dot{\mbox{\boldmath$X$}}=\mbox{\boldmath$f$}(\mbox{\boldmath$X$})
\end{equation}
with the solution $\mbox{\boldmath$X$}(t)=(\mbox{\boldmath$x$},
\dot{\mbox{\boldmath$x$}})$ and its corresponding variational
equation
\begin{equation}
\dot{\mbox{\boldmath$Y$}}(t)=\frac{\partial
\mbox{\boldmath$f$}}{\partial\mbox{\boldmath$X$}}\circ\mbox{\boldmath$Y$}
\end{equation}
with a solution $\mbox{\boldmath$Y$}(t)=(\mbox{\boldmath$\xi$},
\dot{\mbox{\boldmath$\xi$}})$ at time $t$, the maximum LE is given
by
\begin{equation}
\gamma_1 =\lim_{t\rightarrow\infty}~\frac{1}{t}\ln
\frac{|\mbox{\boldmath$Y$}(t)|}{|\mbox{\boldmath$Y$}(0)|}.
\end{equation}
This is called as the variational method. The system is chaotic if
$\gamma_1>0$, otherwise it is regular. This way is rigorous to get
the tangent vector $\mbox{\boldmath$Y$}(t)$, but it is rather
cumbersome to derive Eq. (2) in general. Because of that people
usually use the deviation vector $\Delta \mbox{\boldmath$X$}(t)$
between a reference trajectory $\mbox{\boldmath$X$}(t)$ and a
shadow one $\mbox{\boldmath$\widetilde{X}$}(t)$ as the approximate
tangent vector. This is a less rigorous but still useful
technique, so-called the two-particle method [18,24] or
two-nearby-trajectories method, in which the LE in the expression
of Eq. (3) is to be replaced by
\begin{equation}
\gamma_2 =\lim_{t\rightarrow\infty}\frac{1}{t}\ln
\frac{|\Delta\mbox{\boldmath$X$}(t)|}{|\Delta\mbox{\boldmath$X$}(0)|}.
\end{equation}
Here an initial separation $|\Delta\mbox{\boldmath$X$}(0)|$
relative to $|\mbox{\boldmath$X$}(0)|$  not larger than $10^{-8}$
is viewed as the best choice to guarantee $\Delta
\mbox{\boldmath$X$}$ as a good approximation to
$\mbox{\boldmath$Y$}$ and to avoid the overestimation of LE [24].

As was mentioned in Ref. [23], it is preferable to compute the LE
in the configuration space instead of in the phase space as LEs in
the two spaces are both effective in detecting the long-term
dynamical behavior of orbits. In this case, Eqs. (3) and (4) are,
respectively, modified as follows
\begin{eqnarray}
\lambda_1 &=& \lim_{t\rightarrow\infty}\frac{1}{t}\ln
\frac{|\mbox{\boldmath$\xi$}(t)|}{|\mbox{\boldmath$\xi$}(0)|}, \\
\lambda_2 &=& \lim_{t\rightarrow\infty}\frac{1}{t}\ln
\frac{|\Delta\mbox{\boldmath$x$}(t)|}{|\Delta\mbox{\boldmath$x$}(0)|}.
\end{eqnarray}
For the two particles method, it is necessary to scale the
distance $|\Delta\mbox{\boldmath$x$}(t)|$ down from time to time.
In this way, the shadow trajectory returns to the neighborhood of
the reference one along the deviation vector $\Delta
\mbox{\boldmath$x$}(t)$. The magnitude of $\Delta
\mbox{\boldmath$x$}(t)$ shrinks to the initial distance
$|\Delta\mbox{\boldmath$x$}(0)|$ after each renormalization, and
the velocity deviation vector $\Delta
\dot{\mbox{\boldmath$x$}}(t)$ should be multiplied by the same
factor
$|\Delta\mbox{\boldmath$x$}(0)|/|\Delta\mbox{\boldmath$x$}(t)|$.
Meanwhile, it is also vital to avoid saturation of orbits in a
bounded chaotic region.

The time $t$ and coordinate $\mbox{\boldmath$x$}$ in Eq. (5) or Eq.
(6) are not necessary physical and meaningful in general relativity.
Thus, it is desirable to define covariant LE.
\subsection{LEs in general relativity}

For a given 4-dimensional spacetime with the metric
$ds^2=g_{\mu\nu} dx^{\mu }dx^{\nu}$, first we consider a rigorous
definition of LEs by using a geodesic deviation vector.

\subsubsection{Geodesic deviation vector technique}

In the spacetime, a free particle moves along the geodesic
equation $DU^{\mu}/D\tau=0$, that is to say,
\begin{equation}
\ddot{x}^{\mu}
=-\Gamma^{\mu}_{\alpha\beta}\dot{x}^{\alpha}\dot{x}^{\beta},
\end{equation}
and its geodesic deviation equation is
\begin{equation}
\frac{D^2 \xi^{\mu}}{D\tau^{2}}
=-R^{\mu}_{\alpha\lambda\beta}\xi^{\lambda} U^{\alpha}U^{\beta},
\end{equation}
where $\Gamma^{\mu}_{\alpha\beta}$,
$U^{\alpha}$($=\dot{x}^{\alpha}$) and
$R^{\mu}_{\alpha\lambda\beta}$ stand for the Christoffel symbol,
4-velocity, and the Riemannian curvature tensor, respectively.

From Eq. (5), it is easy to get an invariant definition of LE if
the Riemannian norm and  proper time $\tau$ substitute the
Euclidian norm and  coordinate time $t$, respectively. In the
geodesic deviation vector method (M1) the LE is given in the form
\begin{eqnarray}
\Lambda_{1} &=& \lim_{\tau\rightarrow\infty}\frac{1}{\tau}\ln
\frac{\|\mbox{\boldmath$\xi$}(\tau)\|}{\|\mbox{\boldmath$\xi$}(0)\|}, \\
\|\mbox{\boldmath$\xi$}(\tau)\| &=&
\sqrt{|\mbox{\boldmath$\xi$}\bullet\mbox{\boldmath$\xi$}|}=\sqrt{|g_{\mu\nu}
\xi^{\mu}\xi^{\nu}|}.
\end{eqnarray}
As an illustration, a similar method given in the phase space can be
seen in Ref. [28]. It is easily shown that the Riemannian inner
product $\mbox{\boldmath$\xi$}\bullet\mbox{\boldmath$\xi$}$ is
positive definite because $\mbox{\boldmath$\xi$}(\tau)$ is always
space-like when $\mbox{\boldmath$\xi$}(0)$ is space-like for a given
geodesic flow.

On the other hand, there is not a variational equation like the
form (8) for a non-geodesic flow. However, it is still possible to
derive the variational equation similar to Eq. (2) for the
non-geodesic flow. For example, Nieto \emph{et al.} [29] gave a
relativistic top deviation equation as a generalization of the
geodesic deviation equation for a pair of nearby point particles.
Thus M1 remains valid if the spacetime background is
$ds^2=g_{\mu\nu} dx^{\mu }dx^{\nu}$.

\subsubsection{Two-nearby-orbits method with projection operations}

The derivation of the geodesic deviation equation is usually a
rather hard task. In particular, it is much arduous to obtain the
variational equation of a non-geodesic flow. Thus, it is a good
idea to refine the classical definition by Eq. (6) using an
invariant version of it, introduced by Wu and Huang [23] (method
M2).

Let two particles, an ``observer'' and his ``neighbor'', move on two
nearby trajectories in a curved spacetime. At a coordinate time $t$
the observer is at the point $O$ with coordinate $x^{\alpha}$ and
4-velocity $U^{\alpha}$, and his neighbor reaches the point
$\widetilde{O}$ with coordinate $\widetilde{x}^{\alpha}$. The
deviation vector
\begin{equation}
\Delta \mbox{\boldmath$x$}(t)=\Delta
x^{\alpha}(t)=\widetilde{x}^{\alpha}(t)-x^{\alpha}(t)
\end{equation}
from $O$ to $\widetilde{O}$ is projected to the observer's local
space and the resulting projected vector is $\Delta
x^{\alpha}_{\bot}=h^{\alpha}_{\beta} \Delta x^{\beta}$, where
$h^{\alpha\beta}= g^{\alpha\beta}+U^{\alpha}U^{\beta}$ is the
space projection operator of the observer. The space distance of
the neighbor measured by the observer at time $t$ is
\begin{equation}
\Delta L =  \sqrt{g_{\alpha \beta} \Delta x^{\alpha}_{\bot} \Delta
x^{\beta}_{\bot}}  = \sqrt{h_{\alpha \beta} \Delta x^{\alpha}\Delta
x^{\beta}}.
\end{equation}
Hence we define an invariant LE (M2) as
\begin{equation}
\Lambda_{2}=\lim_{\tau \rightarrow\infty} \frac{1}{\tau} \ln
\frac{\Delta L(\tau)}{\Delta L(0)},
\end{equation}
where the proper time $\tau$ corresponds to the coordinate time $t$
according to the metric. Obviously, this LE is independent of
coordinate transformations [23].

There are detailed implementations of M2 in [23]. Here we
emphasize that the coordinate time is adopted as the common
independent variable for both particles in their equations of
motion and one has to construct an equation for $d\tau /dt$, which
is to be integrated together with the motion equations to get the
proper time of the observer. The reason for doing this  is that
the two particles have their own and different proper times but
the numerical integration demands a common time variable.

\subsubsection{Two-nearby-orbits method without projection operations}

On the other hand, if we integrate directly the equations of motion,
 Eq. (7), for  two slightly distinct initial conditions with the
proper time as an integration variable, we attain the deviation
vector
\begin{equation}
\Delta \mbox{\boldmath$x$}(\tau)=\Delta
x^{\alpha}(\tau)=\widetilde{x}^{\alpha}(\tau)-x^{\alpha}(\tau)
\end{equation}
at the proper time $\tau$ of the observer. Here a difference
between Eq. (11) and Eq. (14) should not be neglected. The
difference is that $\Delta \mbox{\boldmath$x$}$ is a function of
$t$ in Eq. (11), while being a function of $\tau$ in Eq. (14). In
this case, we can give another two-nearby-orbits method (M3) as
follows
\begin{eqnarray}
\Lambda_{3} &=& \lim_{\tau\rightarrow\infty}\frac{1}{\tau}\ln
\frac{\|\Delta\mbox{\boldmath$x$}(\tau)\|}{\|\Delta\mbox{\boldmath$x$}(0)\|}, \\
\|\Delta\mbox{\boldmath$x$}(\tau)\| &=&
\sqrt{|\Delta\mbox{\boldmath$x$}\bullet\Delta\mbox{\boldmath$x$}|}=\sqrt{|g_{\mu\nu}
\Delta x^{\mu}\Delta x^{\nu}|}.
\end{eqnarray}

Next let us study the three methods from the theoretical and
numerical points of view.

\subsection{The reason why M1 and M3 do not
need projection operations}

Obviously, projection operations do not appear in M1. Similarly,
Hartl noted this fact in  his PhD thesis [8]. Now, we give a
discussion on this issue. It has been shown without any difficulty
that $\mbox{\boldmath$\xi$}(\tau)$ is always perpendicular to the
4-velocity $\mbox{\boldmath$U$}(\tau)$ at any (proper) time
provided that $\mbox{\boldmath$\xi$}(0)$ and
$D\mbox{\boldmath$\xi$}(0)/D\tau$ are both perpendicular to
$\mbox{\boldmath$U$}(0)$, respectively [28]. What would happen if
$\mbox{\boldmath$\xi$}(0)\bullet\mbox{\boldmath$U$}(0)\neq 0$ or
$D\mbox{\boldmath$\xi$}(0)/D\tau\bullet\mbox{\boldmath$U$}(0)\neq
0$? Let us explore the question in the following.

Noting that $\mbox{\boldmath$\xi$}\bullet\mbox{\boldmath$U$}$ and
$D\mbox{\boldmath$\xi$}/D\tau\bullet\mbox{\boldmath$U$}$ are
scalars, we can operate a covariant derivative in place of an
ordinary one. The demonstration is
\begin{eqnarray}
\frac{d}{d\tau} (\mbox{\boldmath$\xi$}\bullet\mbox{\boldmath$U$})
&=& \frac{D}{D\tau}
(\mbox{\boldmath$\xi$}\bullet\mbox{\boldmath$U$})
=\frac{D\mbox{\boldmath$\xi$}}{D\tau} \bullet\mbox{\boldmath$U$}
+\mbox{\boldmath$\xi$}\bullet \frac{D\mbox{\boldmath$U$}}{D\tau} \nonumber \\
 &=& \frac{D\mbox{\boldmath$\xi$}}{D\tau} \bullet\mbox{\boldmath$U$},  \\
\frac{d}{d\tau} (\frac{D\mbox{\boldmath$\xi$}}{D\tau}
\bullet\mbox{\boldmath$U$}) &=&
\mbox{\boldmath$U$}\bullet\frac{D^{2}\mbox{\boldmath$\xi$}}{D\tau^{2}}
=-R^{\mu}_{\alpha\lambda\beta}\xi^{\lambda}
U^{\alpha}U^{\beta}U_{\mu} \nonumber \\
&=& -R_{\nu\alpha\lambda\beta} U^{\alpha}U^{\beta} U^{\nu}
\xi^{\lambda}.
\end{eqnarray}
The result of Eq. (18) is obviously identical to zero because it
is both symmetric and antisymmetric with respect to the indicies
$\nu$ and $\alpha$. Consequently, it can be inferred that the
right side of Eq. (17) is a constant $C$, which turns out to be
the value of
$D\mbox{\boldmath$\xi$}(0)/D\tau\bullet\mbox{\boldmath$U$}(0)$. As
was stated above,
$\mbox{\boldmath$\xi$}\bullet\mbox{\boldmath$U$}\equiv 0$ if
$C=0$. Otherwise we have
\begin{equation}
\mbox{\boldmath$\xi$}\bullet\mbox{\boldmath$U$}=C\tau+\widetilde{C},
\end{equation}
where $\widetilde{C}$ is another constant corresponding to the
initial value of
$\mbox{\boldmath$\xi$}\bullet\mbox{\boldmath$U$}$. Now let us
investigate the physical significance of
$|\mbox{\boldmath$\xi$}\bullet\mbox{\boldmath$U$}|$. Obviously it
is no other than the length of the projected vector
$-(\mbox{\boldmath$\xi$}\bullet\mbox{\boldmath$U$})\mbox{\boldmath$U$}$,
the time component of $\mbox{\boldmath$\xi$}$ as measured by the
particle. Equation (19) shows that the evolution of
$\mbox{\boldmath$\xi$}$ along the time direction of the particle
is linear, therefore, the LE in this direction vanishes. In fact,
the result is very natural due to the integral
$U_{\alpha}U^{\alpha}=-1$. Therefore, $\mbox{\boldmath$\xi$}$ and
its projected vector along the space direction of the particle
change at the same rate with its proper time. This implies that it
is not necessary to project $\mbox{\boldmath$\xi$}$ into the space
direction in the calculation of the principal LE. In other words,
it is reasonable to use M1 as the standard definition of the
maximum LE in general relativity.

As to M3, the deviation vector $\Delta \mbox{\boldmath$x$}(\tau)$
in Eq. (14) is viewed as an approximation to the geodesic
deviation vector $\mbox{\boldmath$\xi$}$  by Eq. (8). Therefore,
M3 is nearly the same as M1 as long as the length of
$\mbox{\boldmath$\xi$}$ is kept small enough.

For M2, we use the projected vector $\Delta x^{\alpha}_{\bot}$
rather than  $\Delta \mbox{\boldmath$x$}(t)$ used in Eq. (11) as
an approximation of $\mbox{\boldmath$\xi$}$. $\Delta
\mbox{\boldmath$x$}(t)$ in Eq. (11) is the deviation vector
between the observer and his neighbor at the same coordinate time
$t$, therefore it is necessary for $\Delta \mbox{\boldmath$x$}(t)$
to be projected along the space direction of the observer. On the
other hand, $\Delta \mbox{\boldmath$x$}(t)$ and
$\mbox{\boldmath$U$}$ do not necessarily obey the relation (19)
when $t$ and $\tau$ have large difference. Then in the observer's
in-track direction he may find LE in the form
\begin{equation}
\lambda_{\emph{in-track}}=\lim_{\tau \rightarrow\infty}
\frac{1}{\tau} \ln \frac{\Delta T(\tau)}{\Delta T(0)},
\end{equation}
where $\Delta T (\tau)=|\Delta \mbox{\boldmath$x$}(t) \bullet
\mbox{\boldmath$U$}|$ is the projection of $\Delta
\mbox{\boldmath$x$}(t)$ along the observer's time direction. This
shows that one may find chaos in the observer's time direction if
$\lambda_{\emph{in-track}}$ is used as an index for chaos. This
displays it necessary to employ the projection norm for M2.

We shall further check the validity of the three methods through
practical calculations.

\subsection{Numerical comparisons}

Let us take a Weyl  spacetime as a physical model (see the
appendix), and choose parameters as $E=0.9679$, $L=3.8$ and
$\mathcal{O}=7.012\times10^{-7}$. The initial conditions of two
variables $r$  and $\dot{r}$ are arbitrarily given within their
respective admissible intervals except $\theta=\pi/2$, while
$\dot{\theta}$ is derived from Eq. (A4). First order differential
systems from the geodesic Eqs. (A5) and (A6) are computed with a
7-8 order Runge-Kutta-Fehlberg algorithm of variable time-step
(RKF7(8)). A proper time output interval is 0.1, and the
Poincar\'{e} surface of section is at the plane
$\theta=\frac{\pi}{2}$ ($\dot{\theta}<0$) in Fig. 1. The
intersections of the orbits by this Poincar\'{e} surface of
section describe the global dynamical feature of the system. There
are two regions in the $(r,\dot{r})$ plane. One region with
randomly distributed points is regarded as a chaotic one, while
the other, a regular one consisting of many tori and islands.

Now we quantitatively describe the dynamics of particles moving in
this spacetime by employing the LEs above. For M1, it is necessary
to integrate the dynamical Eqs. (A2)-(A6) numerically with their
geodesic deviation equations together so that we obtain their
solution $(x^{\alpha}, U^{\alpha})$ and their variational solution
$(\mbox{\boldmath$\xi$}, \dot{\mbox{\boldmath$\xi$}})$ in the
forms $\mbox{\boldmath$\xi$}=(\delta t, \delta r, \delta\theta,
\delta\phi)$ and $\dot{\mbox{\boldmath$\xi$}}=(\delta \dot{t},
\delta \dot{r}, \delta\dot{\theta}, \delta\dot{\phi})$. Here the
proper time $\tau$ is chosen as an integration variable. In
addition, the renormalization technique is used if necessary. As
far as M2 is concerned, we numerically trace the trajectories of
the observer and his neighbor with the Eqs. (A7,A8) and the
equations involving $d\tau/d t$ and $d\phi/d t$ with slightly
different initial conditions. The initial conditions of the
observer are the same as the above. As to his neighbor, an initial
separation in the order of $10^{-8}$ is adopted only on the $r$
direction, regarded as the best choice [24], and the other
coordinates remain the same as the observer's except
$\dot{\theta}$. This process also fits M3, but only Eqs. (A2)-(A6)
are integrated numerically.

Choosing many ordered and chaotic orbits as numerical tests, we
found that the three methods M1, M2 and M3 almost give the same
final values of LEs for each orbit. There are only a few
differences on the transition phases of the orbits. For example,
we choose a chaotic orbit with the initial conditions: $t=0$,
$\theta=\pi/2$, $r=9$, $\phi=0$, $\dot{r}=0.025$ and the initial
value of $\dot{\theta}$ derived from Eq. (A4). Let the initial
deviation vector satisfy $\delta t=\delta \theta=\delta
\phi=\delta \dot{r}=\delta\dot{\theta}=0$ and $\delta
r=1/\sqrt{g_{11}}$. As is shown in Fig. 2, M3 is always close to
M1 (that is, M3 is just a good approximation of M1), but M2 is not
consistent with M1 until the stabilizing value of LEs is obtained.
Consequently, the three methods, M1, M2 and M3, have almost the
same final value of LEs.

For conceptual clarity it is of the physical significance to
define LE as a coordinate gauge invariant in general relativity.
For this purpose, there are the above three methods (M1, M2 and
M3) to compute LEs. M1 is more rigorous than M2 or M3. On the
contrary, M2 is more convenient to use than M1, and M3 is the
simplest. In addition, M2 and M3 are easier to treat a
non-geodesic flow as well as a geodesic flow than M1. Numerical
experiments display that M1, M2 and M3 not only can obtain the
same final values but also have small differences in the
transition phase for the calculation of LEs. Considering the
application of convenience, we conclude that M3 should be worth
recommending to calculate LEs in a relativistic gravitational
system. It is useful to accept some advice on a suitable choice of
the initial separation and a renormalization time step introduced
by Tancredi \emph{et al} [24].

Although the LEs have been widely used to distinguish between
regular and chaotic orbits, their stabilizing values obtained
often need rather expensive computational cost in exploring the
global structure of phase space. In view of this, we shall modify
M3 and adopt its corresponding FLI in the following section.

\section{Description of the structure of phase space by using
FLI with two-nearby-trajectories}

\subsection{Fast Lyapunov indicators}
The time interval that is necessary to reach a given value of
either the length of a tangential vector or the angle between two
tangential vectors can be taken as an indicator of stochasticity
of a Newtonian dynamical system. Following this idea,
Froeschl\'{e} \emph{et al.} [21] defined three different FLIs in
1997. However two of these methods need to solve the variational
Eq. (2) for $n$ times when a set of $n$ independent initial
tangential vectors with the same initial conditions are chosen. In
this circumstance, Froeschl\'{e} and Lega [22] improved the FLI as
follows
\begin{equation}
\psi(t) = \ln |\mbox{\boldmath$Y$}(t)|,
\end{equation}
where $\mbox{\boldmath$Y$}$ is a tangential vector from Eq. (2)
and $|\mbox{\boldmath$Y$}(0)|=1$. Given a threshold, the indicator
$\psi$ reaches a value fast for a chaotic orbit, but it would take
a rather long time for an ordered one. Conversely, in the same
time interval the indicator tends to different values for ordered
and chaotic orbits. Namely it grows exponentially with time for
the chaotic orbit, but only algebraically with time in the regular
case. This allows one to distinguish between the two cases. There
is a close relation between the FLI ($\psi$) and the LE [Eq. (3)]:
the FLI divided by time $t$ tends to the LE when the time is
sufficiently large. Besides, overflow of the lengths of tangential
vectors in the case of a chaotic orbit can be avoided because the
integration time is not long enough. This is the reason why the
indicator is classified as a ``fast" method.

Following the idea above and the coordinate gauge invariance, in a
curved spacetime we could easily define an invariant FLI
corresponding to the LE in Eqs. (9) and (15). As far as the
geodesic deviation vector method M1 is concerned, its
corresponding FLI is
\begin{equation}
\Psi(\tau) = \log_{10} \|\mbox{\boldmath$\xi$}(\tau)\|.
\end{equation}
In the light of Eq. (15), we define the FLI with
two-nearby-trajectories as follows:
\begin{equation}
 FLI(\tau) = \log_{10}
\frac{\|\Delta\mbox{\boldmath$x$}(\tau)\|}{\|\Delta\mbox{\boldmath$x$}(0)\|}.
\end{equation}
Next we shall mainly describe the numerical implementation of the
FLI and test its validity in the description of dynamics.

\subsection{The algorithm in detail}
The original FLI proposed by Froeschl\'{e} and Lega [22] requires
to compute the expansion rate of a tangential vector in the
variational equations and it does not need any renormalization. We
now reform it in general relativity into a coordinate invariant
and compute it with the two particles approach. The FLI($\tau$) in
Eq. (23) can not be computed without renormalization because the
distance between the two particles,
$\|\Delta\mbox{\boldmath$x$}\|$, would expand so fast in the case
of chaos as to it could reach the chaotic boundary to cause
saturation.

In our numerical model (see the appendix) we choose
$\|\Delta\mbox{\boldmath$x$}(0)\|=10^{-9}$. We found saturation
when $\|\Delta\mbox{\boldmath$x$}\|=1$, therefore we choose
$\|\Delta\mbox{\boldmath$x$}\|=0.1$ as the critical value to
implement the renormalization. In this way, the number of
renormalization for computing FLI is less than that for LE. This
brings an advantage to guarantee the speed of computation. Let $k
~(k=0,1,2,\cdots)$ be the sequential number of renormalization,
then we calculate the FLI with the following expression
\begin{eqnarray}
FLI_{k}(\tau) &=& -k\cdot(1+\log_{10}
\|\Delta\mbox{\boldmath$x$}(0)\|) \nonumber \\ & & + \log_{10}
\frac{\|\Delta\mbox{\boldmath$x$}(\tau)\|}{\|\Delta\mbox{\boldmath$x$}(0)\|},
\end{eqnarray}
where $\|\Delta\mbox{\boldmath$x$}(0)\|
\leq\|\Delta\mbox{\boldmath$x$}(\tau)\|\leq 0.1$. This technique
depends on the choice of the initial deviation
$\|\Delta\mbox{\boldmath$x$}(0)\|$, and fails when
$\|\Delta\mbox{\boldmath$x$}(0)\|$ is too small or too large. After
many numerical experiments, we found that
$\|\Delta\mbox{\boldmath$x$}(0)\|\approx 10^{-7}-10^{-9}$ works
well.

\subsection{Numerical tests of FLI}

We take the initial separation $\Delta r=10^{-9}$ and choose the
regular orbit $M$ and the chaotic one $N$ in Fig. 1 to test the
sensitivity of the FLI as well as $\Psi$. As is expected, Fig. 3
(colored light gray) displays that there is a drastic difference
of the FLIs between the regular orbit and the chaotic one. For the
ordered orbit, within a time span of $10^{5}$ (equivalent to about
100 periods), the FLI is smaller than one. However, we clearly see
a sharp increase up to 24 of the FLI for the chaotic orbit. In
other words, the length of the geodesic deviation vector
$\mbox{\boldmath$\xi$}$ reaches the value $10^{24}$. As mentioned
above, the vector increases in an algebraic law for quasi-periodic
orbits, but it does with an exponential law for chaotic orbits.
Indeed the chaos of the orbit in the right of Fig. 3 becomes
explicit after a time span $4.5\times 10^{4}$. The FLI is a
cheaper way to distinguish between chaotic and regular orbits and
explore the global qualitative structure of the phase space of a
system.

Another point to emphasize is that the values of FLI are very
analogous to that of $\Psi$ (see black dots in Fig. 3). The values
of FLIs by use of the two methods for the regular case have no
difference in a long time since renormalization is not used. For
the chaotic case we use only two times of renormalization. In
general, it is enough to take small numbers of renormalization to
calculate the FLIs in the orbits with weak chaos. This shows
sufficiently the validity of our FLI with two-nearby-trajectories.

To examine the validity of the FLI further, in the left of Fig. 4 we
have plotted the running maximum of FLIs as a function of the
initial action $r$ for a set of 1657 orbits that are regularly
spaced in the interval [5.77, 22.34] on the axis $\dot{r}=0$ in Fig.
1. Here each orbit is integrated until the time reaches $10^{5}$. As
is described in Fig. 1, the regular region is located in one
interval between $R(8.45, 0)$ and $S(21, 0)$ and another interval
between $R(22.15, 0)$ and the point $(22.34, 0)$ on the straight
line $\dot{r}=0$. Obviously we find that all the FLIs in the two
intervals are not larger than 6, but all the FLIs in the two
intervals [5.77, 8.45] and [21, 22.14] on the straight line
$\dot{r}=0$ are larger. This method displays the set of chaos,
although it does not give the final values of LEs. In a similar way,
the orbits along the straight line $r=9$ in Fig. 1 are also
separated into regular and chaotic intervals (see the right of Fig.
4). These facts show sufficiently that the FLI is rather successful
to distinguish between the regularity and the chaoticity. However,
we must point out a problem that the FLI does not provide more
details about ordered orbits containing quasi-periodic and resonant
periodic orbits. Here we give an illustration in detail.

On the basis of Froeschl\'{e} and Lega [22], the orbit with the
starting point $M(15, 0)$ in Fig. 1 corresponds to a resonance
since the maximum value of FLI arrives at about 0.68 as the
smallest value in the left of Fig. 4. However this is not a
resonant periodic orbit but just a quasi-periodic one. Actually
the  point $I(14.00349, -0.0312004)$ of the Poincar\'{e} map in
the middle of tori is a fixed point. We found that all tori
between the torus $M$ and the point $I$ have almost the same
values of FLI. We cannot say that FLI is appropriate to identify
the resonance categories. As another example, let us focus a
trajectory made up of three little loops or islands surrounding
three invariant points $J_1(9.0975,-0.0281)$,
$J_2(17.13931,0.04913)$ and $J_3(13.94246,-0.08936)$ in Fig. 1,
respectively. An important point to mention lies in the manner
where the three little loops appear on the plot as the same
trajectory. Rather than tracing one loop at a time, successive
points occur at each of the three loops in turn. In this sense,
the three fixed points are inhabited in the same periodic orbit.
The maximum value of FLI we computed for this periodic (resonant)
orbit after a time span of $10^{5}$ turns out to be 1.75 or so.
Obviously the value of FLI for the resonant orbit is larger than
that for the quasi-periodic orbit $M$. In a word, it is difficult
to apply values of FLI to distinguish various regular orbits.

\subsection{Classification of orbits by FLI}
Many experiments above have shown that our FLI is a very sensitive
tool for detecting the regular and chaotic orbits. Next we shall
follow this FLI to scan the global structure of phase space in the
spacetime (A1). This operation is realized by calculating
thousands of orbits in practice. First we fix $\dot{\theta}=0$,
meanwhile we let $r$ run from $r_{min}=5.77$ to $r_{max}=22.34$
with a span of $\Delta r=0.1$. Then, for each given $r$ we can
solve for $\dot{r}$ and find two roots $\dot{r}_{-}$ and
$\dot{r}_{+}$ ($\dot{r}_{-}<0$ and $\dot{r}_{+}>0$) from the Eq.
(A4). Finally we take $\dot{r}$ from $\dot{r}_{-}$ to
$\dot{r}_{+}$ with a sampling interval $\Delta\dot{r}=0.01$. Once
$r$ and $\dot{r}$ are given initially, $\dot{\theta}$ should be
derived by the relation (A4). As is shown in Fig. 5, we have
plotted all the starting points on the $r-\dot{r}$ plane
($\theta=\pi / 2$). According to different values of FLI, we
classify orbits by some dynamical features. This is called as the
description of the global structure of phase-space for the system.

By a comparison of Fig. 5 with Fig. 1, it is clear that the global
dynamical features they depicted are nearly  compatible. As an
emphasis, an advantage to use FLI becomes more apparent because it
applies to systems with an arbitrary  number of dimensions.
Without doubt, it is handy to study the global dynamics of the
spinning compact binaries [10] by virtue of FLI. It is also easier
to probe the variation of the dynamical characteristics as certain
parameters of the system vary.

\section{Summary}

We compared three methods for computing LEs with coordinate gauge
invariance in general relativity. The three methods are the geodesic
deviation vector technique (M1), the two-nearby-orbits method with
projection operations and with coordinate time as the integration
independent variable (M2), and the two-nearby-orbits method without
projection operations and with proper time as the integration
independent variable (M3). The contributions of this work are as
follows.

It is unnecessary to adopt the projection operators for M1 and M3
from the theoretical point of view. Method M3 is the simplest
method for calculating LEs in a relativistic gravitational system
in most cases. As another contribution, we extended FLI to a
coordinate invariant form in relativistic dynamics, and proposed
its algorithm with the two-nearby-trajectories method. This
indicator can rapidly, reliably and accurately distinguish between
ordered and chaotic motions in relativistic astrophysics. Only
when the initial separation and renormalization within a
reasonable amount of time span are chosen appropriately, is this
FLI nearly consistent with the FLI computed with the geodesic
deviation vector technique. However, the former is rather simpler
to use than the latter. As a characteristic, our FLI grows
exponentially with time in the chaotic case, while it grows
algebraically in the case of quasi-periodic trajectories.
Evaluating the different behaviors of this FLI, we successfully
explored the global dynamics of phase space where regions of chaos
and order are clearly identified.

It should be pointed out that a main advantage of the LE and FLI
with two-nearby-trajectories is their easier application in
treating complicated relativistic gravitational systems with high
degrees of freedom, such as multi-body problems and spinning
compact binaries. In the future work, we shall employ an
appropriate numerical integrator and the FLI to find out what
happens in the spinning compact binaries. The global dynamics of
the binary systems will be explored, and the transition from
regular motion to chaos will be considered as some dynamical
parameters of the system vary.

{\appendix \section{Core-shell system and the equations of motion}

A Weyl  spacetime including a non-rotating black hole surrounded by
an axially symmetric shell in Schwarzchild coordinates
$(t,r,\theta,\phi)$ is expressed as [6]
\begin{eqnarray}
ds^2 &=& -(1-\frac{2}{r}) e^{P}dt^{2}+e^{Q-P}
[(1-\frac{2}{r})^{-1}dr^{2} \nonumber \\
& &+r^{2}d\theta^{2}] +e^{-P}r^{2}\sin^{2}\theta d\phi^{2},
\end{eqnarray}
where $P$ and  $Q$ are the functions of $r$ and $\theta$ only.
This system obviously has two integrals, the energy $E$ and the
angular momentum $L$ in the forms
\begin{eqnarray}
\dot{t} &=& Er(r-2)^{-1}e^{-P}, \\
\dot{\phi} &=& Le^{P}r^{-2}\sin^{-2}\theta.
\end{eqnarray}
In addition, the 4-velocity of a particle always satisfies the
constraint
\begin{eqnarray}
U_{\alpha}U^{\alpha} &=& -(1-\frac{2}{r}) e^{P}\dot{t}^{2}+e^{Q-P}
[(1-\frac{2}{r})^{-1}\dot{r}^{2} \nonumber \\
& &+r^{2}\dot{\theta}^{2}] +e^{-P}r^{2}\sin^{2}\theta
\dot{\phi}^{2}=-1.
\end{eqnarray}
If a fourth integral holds, the system is integrable. Otherwise, it
is possible to yield chaos. However, it is not easy to deduce
whether the fourth integral exists or not. To study the dynamics of
a geodesic particle in this system further, we need the following
geodesic equations
\begin{eqnarray}
\ddot{r} &=& [\frac{1}{r(r-2)}+ \frac{1}{2}(\frac{\partial
P}{\partial r}-\frac{\partial Q}{\partial r})]\dot{r}^{2}-
(\frac{\partial Q}{\partial \theta}-\frac{\partial P}
{\partial \theta})\dot{r}\dot{\theta} \nonumber \\
& & +(r-2)[1+\frac{r}{2}(\frac{\partial Q}{\partial r}
-\frac{\partial P}{\partial r})]\dot{\theta}^{2}+f_1(r,\theta), \\
\ddot{\theta} &=& \frac{1}{2r(r-2)} (\frac{\partial Q}{\partial
\theta}-\frac{\partial P}{\partial \theta})\dot{r}^{2}
-(\frac{2}{r}+\frac{\partial Q}{\partial r}-\frac{\partial
P}{\partial r})
\dot{r}\dot{\theta} \nonumber \\
& & -\frac{1}{2}(\frac{\partial Q}{\partial \theta}
-\frac{\partial P}{\partial
\theta})\dot{\theta}^{2}+f_2(r,\theta),
\end{eqnarray}
where
\begin{eqnarray}
f_1(r,\theta) &=& -\frac{1}{2}(1-\frac{2}{r})e^{2P-Q}
[\frac{2}{r^2}+(1-\frac{2}{r})\frac{\partial P}{\partial r}]\dot{t}^{2} \nonumber \\
& & +\frac{1}{2}(r-2)e^{-Q}\sin^{2}\theta \cdot
(2-r\frac{\partial P}{\partial r})\dot{\phi}^{2}, \nonumber \\
f_2(r,\theta) &=& -\frac{1}{2r^2}(1-\frac{2}{r})e^{2P-Q}
\frac{\partial P}{\partial \theta} \cdot\dot{t}^{2}+\frac{1}{2}e^{-Q} \nonumber \\
& & \cdot\sin\theta \cdot(-\frac{\partial P}{\partial
\theta}\sin\theta+2\cos\theta)\dot{\phi}^{2}. \nonumber
\end{eqnarray}
We now have to do the tedious derivation of the geodesic deviation
equations. The result is too complex and will not be written here.
If we take $a'$ as the derivative of $a$ with respect to coordinate
time $t$, the above geodesic equations are readjusted as follows
\begin{eqnarray}
r'' &=& \frac{\ddot{r}}{\dot{t}^{2}}-\frac{r'}{\dot{t}}\frac{d\dot{t}}{dt}, \\
\theta'' &=& \frac{\ddot{\theta}}{\dot{t}^{2}}
-\frac{\theta'}{\dot{t}}\frac{d\dot{t}}{dt}, \\
\frac{d\dot{t}}{dt} &=& -Er(r-2)^{-1}e^{-P}(\frac{\partial
P}{\partial r}r'
+\frac{\partial P}{\partial \theta}\theta') \nonumber \\
& & -2E(r-2)^{-2}e^{-P}r'. \nonumber
\end{eqnarray}
Here $\dot{r}$ and $\dot{\theta}$ in Eqs. (A5) and (A6) should be
changed into $\dot{r}=r'\dot{t}$ and $\dot{\theta}=\theta'\dot{t}$
respectively, while $\dot{t}$ and $\dot{\phi}$ remain the original
expressions of Eqs. (A2) and (A3) because they do not contain
coordinate time $t$ explicitly.

Given two metric functions $P$ and $Q$, the spacetime (A1) is
determined at once. Now we reexamine the full relativistic
core-shell configuration involving a Schwarzschild black hole plus a
pure octopole shell studied by Vieira \& Letelier [6]. In this
model, we have
\begin{eqnarray}
P(u,v) &=& \frac{1}{5} \mathcal{O}uv(5u^{2}-3)(5v^{2}-3),
\nonumber \\
Q(u,v) &=& -\frac{2}{5}\mathcal{O}v[5(3u^2-1)(1-v^2)-4] \nonumber \\
& & +\frac{3}{100}\mathcal{O}^2[-25u^6(1-v^2) \nonumber \\
& & \cdot(5v^2+2v-1)(5v^2-2v-1) \nonumber \\
& & +15u^4(1-v^2)(65v^4-40v^2+3) \nonumber \\
& & -3u^2(1-v^2)(25v^2-3)(5v^2-3) \nonumber \\
& & -v^2(25v^4-45v^2+27)] \nonumber,
\end{eqnarray}
where $u=r-1$, $v=\cos\theta$ and $\mathcal{O}$ is an octopolar
parameter.

}

\begin{acknowledgments}
We would like to thank Professor George Contopoulos for his
comments and suggestions. Special thanks go to him for his help in
the improvement of the language. We are also grateful to Dr.
Xiao-Sheng Wan of Nanjing University for his effort of checking
our numerical results. This research is supported by the Natural
Science Foundation of China under Contract Nos. 10233020, 10303001
and 10563001, Jiangsu Planned Projects for Postdoctoral Research
Funds and Science Foundation of Jiangxi Education Bureau.
\end{acknowledgments}

\begin{figure}[h]
\includegraphics[scale=0.8]{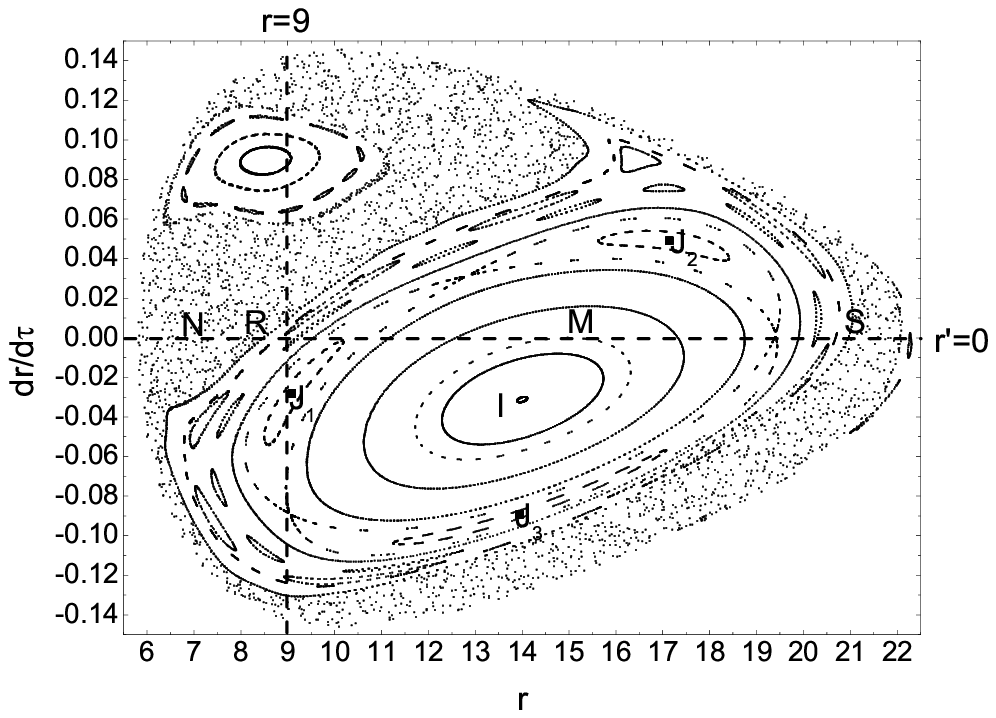}
\caption{\label{fig:epsart}Poincar\'{e} surface of section at the
plane $\theta=\frac{\pi}{2}$ ($\dot{\theta}<0$) for the full
relativistic core-shell configuration consisting of a
Schwarzschild black hole and a purely  octopolar shell with
parameters $E=0.9679$, $L=3.8$ and $\mathcal{O}=7.012\times
10^{-7}$.} \label{fig1}
\end{figure}

\begin{figure*}[ht]
\includegraphics[scale=0.8]{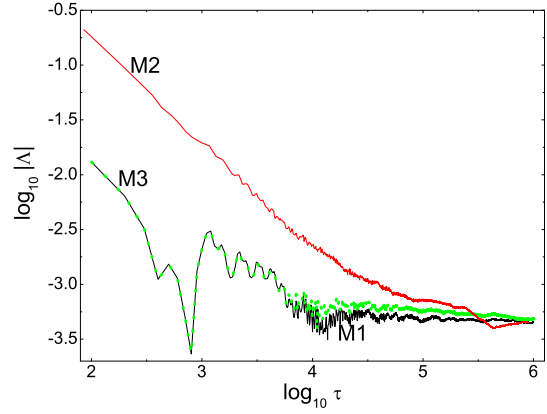}
\caption{LEs for a chaotic orbit with the initial conditions $r=9$
and $\dot{r}=0.025$. The three methods, M1, M2 and M3, give almost
the same final value of LEs.} \label{fig2}
\end{figure*}

\begin{figure*}[ht]
\includegraphics[scale=0.7]{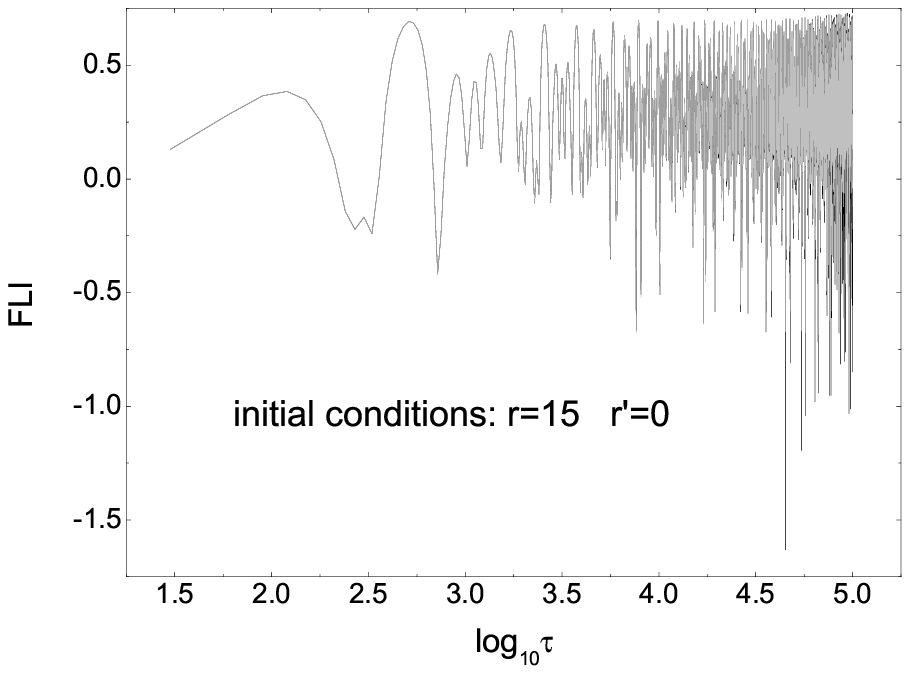}
\includegraphics[scale=0.7]{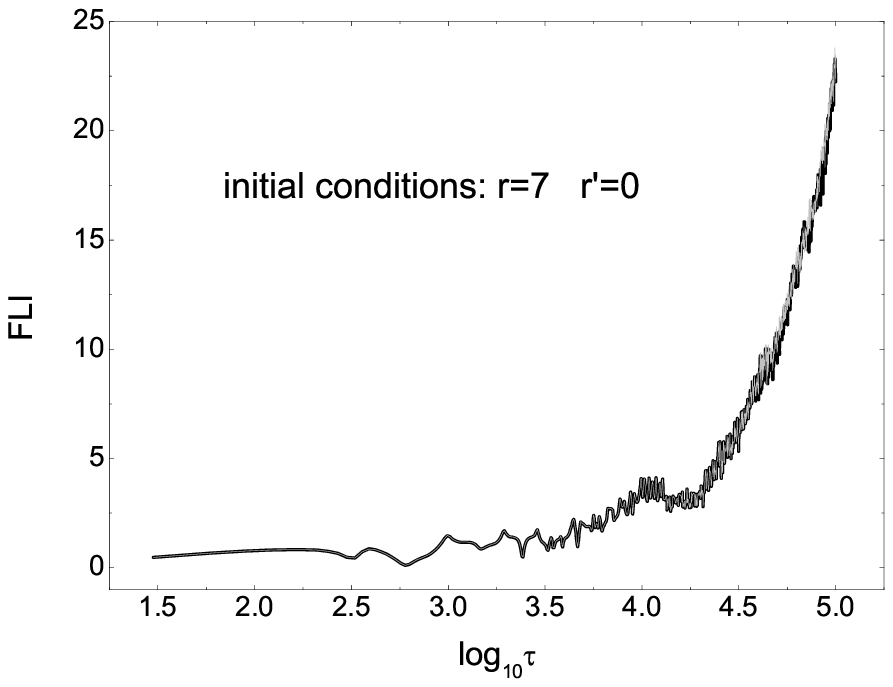}
\caption{The relation of FLIs  with proper time by use of the
index (23) for the regular orbit $M$ in the left (colored light
gray). The black dots correspond to the result obtained from Eq.
(22). The right is the same as the left but for the chaotic orbit
$N$ in Fig. 1.} \label{fig3}
\end{figure*}

\begin{figure*}[ht]
\includegraphics[scale=0.7]{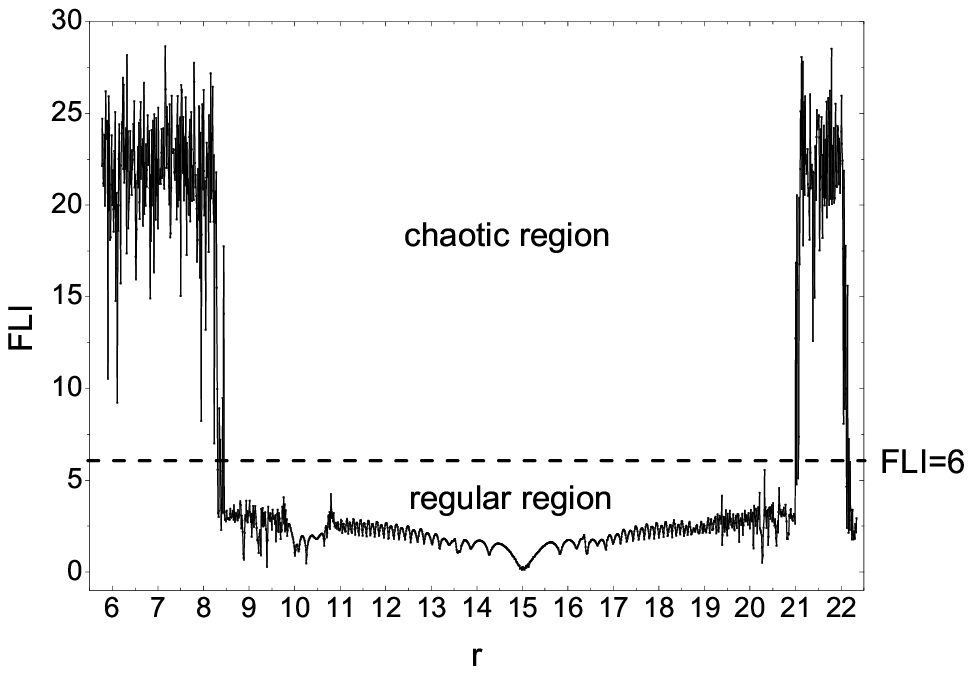}
\includegraphics[scale=0.7]{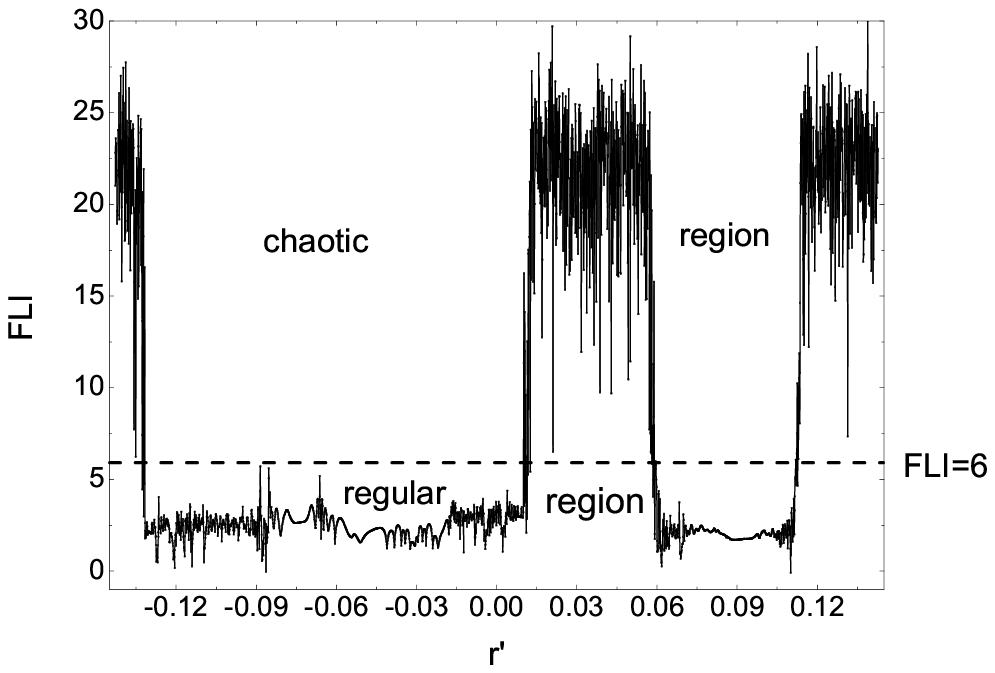}
\caption{Left: the evolution of the FLIs with the initial values
of $r$ ranging from the interval $[5.77,22.34]$ when the indicator
(23) is used and $\dot{r}=0$ is fixed at the initial time. Each
orbit is integrated numerically till $\tau$ reaches $10^{5}$. Two
types of orbits consisting of regular and chaotic orbits are
obtained according to distinct values of FLIs. Right: the same as
the left, but $r=9$ is fixed first, and $\dot{r}$ takes values
from the interval $[-0.1428,0.1428]$.} \label{fig4}
\end{figure*}

\begin{figure}[ht]
\includegraphics[scale=0.8]{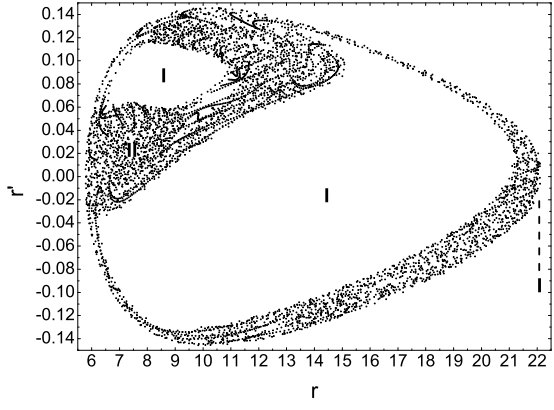}
\caption{Regions of different values of the FLIs on  the plane
$\theta=\frac{\pi}{2}$. Initial conditions are colored white if
their FLIs$\leq$6, and black if FLIs$>$6, which are corresponded
to the regular region I, and chaotic region II, respectively.
Here, all intersection points of an orbit with the plane should be
regarded to have the same final value of the FLI.} \label{fig5}
\end{figure}

\end{document}